\documentclass[11pt,letterpaper,twoside]{book}

\begin{document}

\title{\bf{ Matter-Antimatter Asymmetry in the Universe and an Arrow for Time}}

\author{R. D. Peccei\\
Department of Physics and Astronomy\\
University of California at Los Angeles\\
Los Angeles, California, 90095}
\maketitle
\vspace{1cm}

\section{Introduction}
In the Big Bang model, the Universe starts out as very hot with all
particle species in equilibrium. In these circumstances, creation and
annihilation of matter and antimatter is equally probable. As the
Universe expands it cools down and when the temperature of the Universe
goes below a temperature of $O(T \sim 1{\rm{GeV}})$ the inverse
annihilation process
$2\gamma \to p+\bar{p}$ is energetically blocked. Hence, around this
temperature, the direct process $  p + \bar{p} \to 2 \gamma$  begins to
turn all protons and antiprotons into photons.
This process is very efficient and, if one starts in a matter -
antimatter symmetric Universe, one expects the Universe now to have very
little matter (or antimatter) left. In detail, \cite{Turner} one
predicts a ratio of the baryon to photon densities now of order
\begin{equation}
  \eta_{\rm{Matter}} = \frac{n_B}{n_{\gamma}} \equiv \eta_{\rm{
Antimatter}} \simeq 10^{-18}.
\end{equation}

In fact, the Universe appears to be matter dominated, with little or no
antimatter \cite{CDRG}: $\eta_{\rm{Matter}}>> \eta_{\rm{Antimatter}}$
Furthermore, one finds  that the Universe now has much more matter than
expected. The ratio $\eta $  of baryons to photons is very well
determined  by data from the cosmic background anisotropy measured by
WMAP \cite{WMAP} and  by Big Bang Nucleosynthesis (BBN): \cite{BBN}
\begin{equation}
 \eta=(6.097 \pm 0.206) \times 10^{-10}.
\end{equation}
This result for $\eta $  makes sense only if this parameter is not a
measure of $ n_B /n_{\gamma}$,  but rather is a measure of some
primordial asymmetry:
\begin{equation}
 \eta \equiv \frac{ n_{\rm{Matter}}-
n_{\rm{Antimatter}}}{ n_{\gamma}}|_{\rm{Primordial}}.
\end{equation}
Then, the question becomes one of whether one can explain the origin of
this primordial matter-antimatter asymmetry from the laws of physics?

\section{The Sakharov Conditions}

Remarkably, in 1967 Andrei Sakharov \cite{Sakharov} identified the
physical conditions needed for a matter-antimatter asymmetry to be
established in the Universe. These three conditions are:

\noindent i) {\it The theory must violate fermion number.} To
establish a matter-antimatter asymmetry there must exist  processes
where matter can  disappear, otherwise   $n_{\rm{Matter}} -
n_{\rm{Antimatter}}$ is a constant and $\eta$ is set by the initial
condition assumed for the Universe, and thus is not determined by
physical processes.

\noindent ii) { \it The laws of nature must violate C and CP.} If either C or CP is a good symmetry, the rates of decay of matter and
antimatter are the same [$  \Gamma_{\rm{Matter}} = \Gamma_{\rm{Antimatter}}$]. As
a result no asymmetry can result.

\noindent iii) {\it The fermion number violating processes must be out of
equilibrium in the Universe.}  In equilibrium processes that
create baryon number [ e.g. $e^+ +\pi^o  \to p$] or destroy baryon
number [e.g. $p \to e^+ + \pi^o$] are equally effective, so no
asymmetry can be established.

I would like to discuss in some detail the first two requirements above, as there are some subtleties involved.

\section{Fermion Number Violation}

In the Standard Model (SM), classically, both Baryon number (B) and
Lepton number (L ) are conserved. However, because the fermions are in
chiral asymmetric representations of $SU(2)\times U(1)$,  both the B-
and L-symmetry currents have a chiral anomaly. \cite{ABJ} These
anomalies are the same for both currents, so that the (B-L) combination
survives as an exact symmetry at the quantum level, while for the (B+L)
current one finds
\begin{equation}
\partial_{\mu} J^{\mu}_{\rm{B+L}}=\frac{N_g}{16\pi^2}[g_2^2W_{a\mu\nu}
\tilde{W}^{\mu\nu}_a +g_1^2 Y_{\mu\nu}\tilde{Y}^{\mu\nu}],
\end{equation}
where $N_g$ is the number of generations. Even though the (B+L)-current is not conserved,  't Hooft \cite{'tH}
showed that the actual rate for (B+L)-violating processes at T=0  is
tiny. However,  sphaleron processes \cite{KM} in the thermal bath
provided by an expanding Universe can substantially alter this rate,
and at temperatures near where the electroweak phase transition takes place
(B+L)-violating  processes can occur at a significant rate. This
observation by Kuzmim, Rubakov and Shaposhnikov \cite{KRS} has
significant implications for the matter-antimatter asymmetry in the
Universe.

Before considering these implications in detail, let me comment briefly on the
physics behind this result. The processes at T=0 which contribute to
(B+L)-violation involve gauge field configurations which are
topologically distinct from each other at $t=-\infty$ and $ t=+\infty$.
\cite{'tH} Effectively, these processes require quantum tunneling
between the EW vacuum states and, because of this, the rate for
(B+L)-violating processes is suppressed exponentially. In the thermal
bath of an expanding Universe, on the other hand, the transitions
between EW vacuum states can occur through thermal fluctuations. \cite{KRS}
Furthermore the energy barrier between the vacuum states, which is given
by the Boltzman factor \cite{KM} $\rm{exp}[E_{\rm{sph}}(T)/T]$ decreases with
T, since the sphaleron energy $E_{\rm{sph}}(T) $ vanishes at the
electroweak phase transition, leading to ubiquitous (B+L)-violation in
the early Universe.

One can show that, at temperatures above the electroweak phase
transition, the rate for (B+L)-violating processes remains
significant.\cite{Arnold,B} Indeed,  this rate scales as
\begin{equation}
 \Gamma_{\rm {(B+L)violation}} \sim \alpha^5 ln \alpha ~T,
\end{equation}
and is faster than the Universe's  expansion rate $H\sim T^2/M_P$ for
\begin{equation}
  T_{\rm{EW}} \sim 10^2 {\rm{GeV }} \leq   T\leq    T_{\rm{max}}
\sim 10^{12} {\rm{GeV}}.
\end{equation} 
This has 2 consequences:

\noindent      i) Any primordial (B+L) asymmetry generated before
$T_{\rm{max }}$ gets erased, as (B+L)-violating process go into
equilibrium.

\noindent      ii) As a result of sphaleron re-processing, a primordial
(B - L) asymmetry generates both a B and an L asymmetry. Thus $\eta$ is
naturally associated with a primordial (B - L) asymmetry.

\section{An Arrow for Time}

Because of the CPT Theorem, \cite{CPT} CP-violation is equivalent to a
violation of Time reversal (T). Because CP-violation is a necessary
ingredient for generating a primordial matter-antimatter asymmetry, our
own existence is predicated on having an {\bf{arrow for
time}}. Typically, under CP, operators are replaced by their Hermitian
adjoints. \cite{RDPS} Thus, for example, under CP a $W^+$ field goes
into:
\begin{equation}
W^{+\mu}(\vec{x}, t)\to \eta(\mu) W^{ -\mu} (-\vec{x}, t),
\end{equation}
where  $\eta(0)=-1; \eta(i)=1$. Thus, schematically, under CP:
\begin{equation}
 O(\vec{x}, t) \to O^{\dagger}(-\vec{x}, t)
\end{equation}
However, because  Lagrangians are Hermitian structures, if a Lagrangian
contains an operator O it also contains $O^{\dagger}$ and one has the
structure:
\begin{equation}
  L= a O + a^* O^{\dagger},
\end{equation}
 where a is a c-number. Hence, one sees that, under CP,
\begin{equation}  
                        L \to L ~~{\rm {only~ if}}~~ a= a^*.
\end{equation}
 Thus CP-violation and T-violation require having {\bf {complex
structures}} in the theory.

The origin of this complexity is not understood, but an intriguing idea
is that perhaps it might be the result of compactification of a
higher-dimensional theory to 4-dimensional space-time. Indeed, it was
noted long ago by Dine, Leigh and  MacIntire \cite{DLM} and by Choi, Kaplan
and Nelson \cite{CKN} that in 10-dimensional heterotic string theory,
ordinary 4-dimensional CP is an exact symmetry, being the product of a
gauge transformation and a 10-dimensional Lorentz transformation. 
This is easily understood by noting that in 3-space dimensions Parity
can be viewed as a reflection in one dimension and a rotation by $\pi$
in the plane orthogonal to the axis that was reflected. That is,
\begin{equation}
\{\vec{x} \to -\vec{x}\} \equiv \{x_1 \to - x_1 ~\rm{and}~
R_{x_2x_3}[\pi] \}.
\end{equation}
In 4-space dimensions, however, one can invert all the coordinates
simply by performing two rotations:
\begin{equation}
\{\vec{x} \to -\vec{x}\} \equiv \{ R_{x_1x_2}[\pi] R_{x_3x_4}[\pi] \}.
\end{equation}
So it is clear that Parity in 4-dimensional space time can be part of a
Lorentz transformation in ten dimensions.

Charge conjugation interchanges fermions with anti-fermions. Because in
the heterotic string theory the fermions and anti-fermions transform
both as members of the adjoint representation of the local $E_8$
symmetry of the theory, charge conjugation is equivalent to a gauge
rotation. Hence, in these theories, 4-dimensional CP is an exact symmetry,
being a product of a Lorentz rotation and a gauge transformation. So, if
our 4-dimensional world originates from theories of this sort, where
4-dimensional CP is a good symmetry in the higher dimensional theory,
the complexity which gives rise to CP-violation arises as a result of
the compactification of the extra dimensions and may well be
characterized by a simple geometrical phase.

Regardless of whether this speculation has merit or not, in 4-dimensions
CP violation is naturally associated with the presence of scalar fields.
This is because a theory involving only fermions and gauge fields is
CP-conserving, up to $\theta$-terms, since the gauge couplings are real
and the gauge fields sit in (real) adjoint representations. Furthermore,
at least in the Standard Model (SM), the $\theta$-terms do not appear to
be effective sources for CP violation. The weak $\theta$-angle can
be rotated to zero because $SU(2)$ is a chiral  theory and the strong
$\theta$-angle must be very small [$\theta_s \leq 10^{-10}$] to satisfy
the stringent bound that exists on the electric dipole moment of the
neutron.

If there are no elementary scalars, one could imagine that CP is broken
spontaneously by the dynamical formation of  some complex fermion
condensates : $ <\bar{f} f> \sim e^{i\delta}$. However, it is difficult
to reconcile spontaneous CP-breaking with cosmology.  As Zeldovich,
Kobzarev and Okun \cite{ZKO} pointed out, if CP is broken spontaneously,
domains of different CP form in the Universe, separated by walls which
dissipate very slowly as the Universe cools as it expands.  Typically,
the energy density in the domain walls only decreases linearly with
temperature $\rho \sim \sigma T $, where $\sigma$ is the surface energy
density and is directly related to the scale of the CP-violating
condensates.  If this scale, say, is of order the Fermi scale $v_F
\simeq$ 250 GeV so that $\sigma \sim  v_F^3$, the energy density in the
domain walls now would vastly exceed the closure density for the
Universe.

Because of these considerations, it is natural to assume that the
CP-violating phenomena observed experimentally are due to the presence
of a scalar sector. Indeed, all data in K and B decays are perfectly
consistent \cite{Burchat} with CP violation being due to the CKM
paradigm, \cite{CKM} where CP violation originates from complex Yukawa
couplings. However, it is unlikely that the CKM phase $\gamma$ is directly
related to the CP phase associated with the generation of the observed
matter-antimatter asymmetry in the Universe. Let me turn to a more
detailed discussion of this matter next.

\section{Scenarios for Generating $\eta$}

In the years since Sakharov's seminal paper many scenarios have been
proposed for generating the matter-antimatter asymmetry in the Universe.
Here I will describe briefly the elements of four of these scenarios,
focusing on their viability and their connection to some other physics.

\subsection{GUT-scale Baryogenesis}

Grand Unified Theories (GUTS) \cite{GUT} provided the first
realistic models for baryogenesis. \cite{Yoshimura}
In GUTs  the SM forces are unified into a larger group G which contains
the $SU(3)\times SU(2) \times U(1)$ symmetry group of the SM. As a
result, quarks and leptons are in same representations, leading
naturally to nucleon instability. The proton lifetime, however, is very
long because proton decay is mediated by the very heavy force carriers
$M_X~\sim 10^{16} \rm{ GeV}$ which lead to the unification of forces.

In GUTs  all 3  Sakharov's  conditions are satisfied.  First of all, as
remarked already above, the heavy GUT states couple to channels with
different baryon number. For instance, the charge 4/3 gauge boson X of
$SU(5)$ couples to both two u-quarks and to an anti d-quark and a
positron. Hence X-decays violate baryon number:
\begin{equation}
X \to u u ~ ( B = 2/3) ~~; ~~  X \to e^+ \bar{d} ~   (B = - 1/3).
\end{equation}
 Second,  in GUTs  the rates for  these decays for particles and
anti-particles are not the same, since there are CP-violating phases in
these theories. Hence, for example:
\begin{equation}
  \Gamma_X(X \to u u ) \neq  \Gamma_{\bar{X}}(\bar{X} \to \bar{u}\bar{ u} ).
\end{equation}
Third, in the early Universe, at temperatures of order the mass of the
heavy GUT states $T\sim M_X$, the baryon number violating decays are out
of equilibrium since their decay rates are slow compared to the
Universe's expansion rate.

Although it is possible in GUTs to obtain $\eta \simeq 6 \times
10^{-10}$, \cite{Langacker} this is difficult to do since, in general,
the rate difference $  \Gamma_X(X \to u u ) -
\Gamma_{\bar{X}}(\bar{X} \to \bar{u}\bar{u})$  which drives $\eta$ involves highly
suppressed CP-violating processes. However, the main difficulty of GUT
Baryogenesis is that it creates a (B+L)- asymmetry. \footnote{There are
further troubles for GUT baryogenesis due to the fact that the matter
-antimatter asymmetry is created at a very high temperature. Such
temperatures are difficult to achieve in the re-heating that
occurs after inflation.} As pointed out by Weinberg \cite{Weinberg} and by
Wilczek and Zee \cite{WZ}, the dominant dimension six SM operators that
violate B, actually conserve (B-L). For example, the operator
\begin{equation}
   O_1 = \epsilon_{\alpha \beta \gamma} \epsilon_{ij} (d^T_{\alpha R} C
u_{\beta R})(Q^T_{\gamma i L} C L_{jL})
\end{equation}
has B-L=0. Thus GUT baryogenesis really produces a (B+L)- asymmetry.
However, as we discussed earlier, sphaleron processes
will erase any primordial (B+L)-asymmetry as the Universe cools below $T
\sim 10^{ 12}$ GeV. So, if GUT baryogenesis only produces a
(B+L)-asymmetry, this cannot be the origin of the observed $\eta$!

\subsection{Electroweak Baryogenesis}

A very interesting alternative to GUT baryogenesis is to imagine that
baryogenesis occurs at the EW phase transition. \cite{Shap} Unfortunately,
this scenario also has significant difficulties. The advantages of
electroweak baryogenesis, if it were possible, is that if the
electroweak phase transition is a first order transition, then fermion
number violating processes would naturally be out of equilibrium.
Furthermore, the resulting asymmetry (which is really a (B+L)-asymmetry)
would be driven by CP-violating processes at the weak scale and so the
asymmetry would be proportional to the CKM phase: $\eta \sim \gamma$.

Unfortunately, for the SM the dynamical conditions for electroweak
baryogenesis do not work out in detail. First of all, the CP-violating
processes which contribute to $\eta$  are very suppressed by  GIM
factors and the smallness of the Jarlskog determinant \cite{S} which
result in a calculated value  for $ \eta << 10^{-10}$.
Second, because the mass of the Higgs boson in the SM is known
experimentally to be rather heavy ($M_H >114$ GeV \cite{LEP}), it follows
that the electroweak phase transition is not a very strong first order
transition. This also has dire consequences, since in this case it turns
out that (B+L)-violating processes are still fast enough to be in
equilibrium after the phase transition and serve to erase the produced
asymmetry. In detail, for the asymmetry created at the electroweak phase
transition not to be  erased after the transition, one needs
\begin{equation} 
 \Gamma_{\rm{(B+L)violation}} (<\Phi^*>) <  H,
\end{equation}
where $<\Phi^*>$ is the Higgs VEV after the transition. This inequality
holds if $<\Phi^*>/T^* \geq 1$. \cite{jump} In more physical terms, the condition on the jump of the Higgs vacuum expectation value can be translated into an upper bound on the Higgs mass
$M_H< 45$ GeV \cite{erase} which, unfortunately, is in contradiction with experiment.

It is possible to ameliorate both these troubles in supersymmetric
theories, \cite{Quiros} but the resulting scenarios  for the
supersymmetric matter are quite constrained and $\eta$ is no longer
directly related to the CKM phase $\gamma$. In fact, realistic
supersymmetric models of electroweak baryogenesis in general need to
introduce additional singlet fields to produce the asymmetry.
\cite{singlet} Other extensions of the SM also can produce the desired
asymmetry, but they all need to introduce additional unwarranted
assumptions. For instance, a recent paper by Fromme, Huber and Seniuch,
\cite{FHS} in a 2 Higgs model, finds a strong first order transition for
heavy extra Higgs states [$m_H > 300$ GeV]. In this case, the  CP phase
that drives $ \eta$ is that of  the dimension two mass term which connects
the two Higgs fields
\begin{equation}
 V_{\rm{break}} = \mu_{12}e^ {i\delta} (\Phi_1^TC \Phi_2) +
\mu_{12}e^ {-i\delta} (\Phi_1^TC \Phi_2)^{\dagger}
\end{equation}
 This term breaks CP and  a discrete symmetry which protects this theory
from flavor changing neutral currents (FCNC). However, there appears to be no motivation for introducing this term, except for helping generate a large enough $\eta$.
Furthermore, to generate a large enough asymmetry one needs parameters
which produce an electric dipole moment for the neutron and the electron
very close to the present bounds, and the strong first order phase
transition occurs near the perturbative breakdown of the model.

\subsection{Baryogenesis through Leptogenesis}

A much more promising scenario for baryogenesis is the leptogenesis
scenario suggested by Fukugita and Yanagida \cite{FY} twenty years ago.
\cite{BPY} In this scenario, a primordial lepton-antilepton asymmetry is
generated from the out of equilibrium decays of heavy Majorana neutrinos
into light neutrinos and Higgs bosons   [$N\to \ell \Phi; N \to
\ell^{\dagger}\Phi^{\dagger}$]. Because the     (B+L)- current is
anomalous, this lepton number asymmetry, is transmuted through sphaleron
processes \cite{KRS} into a baryon number asymmetry. After a simple
calculation, \cite{HT} one finds in the SM :
\begin{equation}
 \eta = -\frac{[8N_g+4]}{[22N_g+13]} \eta_L\simeq -0.35 \eta_L,
\end{equation} 
where in the above $\eta_L$ is the
leptonic asymmetry produced in the heavy neutrino decays.

The nicest feature of this scenario is that baryogenesis is related to
another physical phenomena, since the heavy neutrino states N with masses $M_N>>v_F$ whose
decays produce the asymmetry explain also, via the seesaw mechanism,
\cite {seesaw} why neutrino masses are tiny: 
\begin{equation}
m_{\nu}\sim \frac{ m_{\ell}^2}{ M_N}. 
\end{equation}
Thus, $\eta_L$ depends on properties of the light neutrino spectrum and
so does the matter-antimatter asymmetry $\eta$.

To produce the desired baryon asymmetry, one needs a lepton number
asymmetry of order $ \eta_L \simeq 1.7\times 10^{ -9}$. Now,
\begin{equation}
\eta_L \simeq \frac{7 \epsilon \kappa}{ g^*}.
\end{equation}
where $\epsilon$ is a measure of the CP-asymmetry in the decay of the
heavy neutrino N, $\kappa$ takes into account of a  possible washout of
the asymmetry after it is established, and $g^*\sim 100$ is the number of
effective degrees of freedom at the temperature where the asymmetry is
produced: $T\sim M_N$.

The parameter $\epsilon$ vanishes if there is no CP violation in N
decays and is
related to the neutrino spectrum:
\begin{equation}
\epsilon=-\frac{3M_1}{16\pi^2v_F^2}\frac{\rm{ Im}~(
\Gamma^*M_{\nu}\Gamma^{\dagger}) }{ (\Gamma \Gamma^{\dagger})_{11}}.
\end{equation}
Here $M_{\nu}$ is the light nutrino mass matrix and $\Gamma$ is the Yukawa coupling matrix of the heavy neutrinos to the light neutrinos and the Higgs boson. Here we have assumed the case of hierarchical heavy neutrinos, where the
asymmetry is generated by the decay of the lightest such neutrino (whose
mass we have denoted by $M_1$).
 For $T \sim M_1\sim 10^{10}$ GeV, one needs very light neutrino masses
($m_{\nu} <$ eV) to obtain the typical parameters needed for the CP
asymmetry [$\epsilon \sim 10^{-6}$] and the washout $[\kappa \sim
10^{-2}]$ to get $\eta_L \sim 2 \times  10^{ -9}$. 
One of the nice results that emerges is that, if the light neutrino
masses lie in the range $10^{-3} \rm{eV}\leq m_{\nu}\leq$ eV, then the
washout factor $\kappa$ is independent of  the initial abundance of the
heavy neutrinos and/or any pre-existing lepton asymmetry. \cite{BDP}

Because the CP-asymmetry parameter $\epsilon$ cannot be to small for
leptogenesis to work, this provides a lower bound on $ M_1$. \cite {DI}
The analysis of Buchm\"uller, Di Bari and Pl\"umacher \cite{BDP} yields the
bound $M_1> 2\times  10^9$ GeV.   \cite{BDP}
It turns out that leptogenesis also provides an upper bound on the light
neutrinos masses, since the washout rate increases proportionally to the
sum of the light neutrino masses squared, $W \sim \Sigma_i  m^2_i
$. Buchm\"uller {\it{et al}} \cite{BDP2} find the bound:
\begin{equation}
m_i < 0.1 ~\rm{eV}.
\end{equation} 
Obviously, the realization that leptogenesis occurs  as the result of the
out-of-equilibrium decays of heavy neutrinos with masses in the $10^{10
}$ GeV range, and that these heavy neutrinos are associated with a sub
eV light neutrino spectrum is very encouraging!

Leptogenesis is a triumph for neutrino physics. Indeed, if $\eta$ is
generated this way, we owe our own existence to the CP phases in the
neutrino sector! However, the fact that leptogenesis occurred at
temperatures T of order $T\sim M_1> 2\times 10^9 $ GeV  has significant
import for supersymmetric  theories.  In particular, in supergravity
theories if the reheating temperature after inflation $T_R$ is too high,
one overproduces gravitinos. This has  catastrophic consequences for the
evolution of the Universe, since the decay products of gravitinos end up
by destroying the light elements produced in Big Bang Nucleosynthesis.
To avoid  these troubles, one  requires that the reheating temperature
be bounded by $T_R < 10^7 $ GeV. \cite{Kawasaki} However, leptogenesis
argues that $ T_R > M_1 > 2\times  10^9$ GeV!

There are solutions to this, so-called, gravitino problem, but these in
general alter the  "normal" SUSY expectations coming from supergravity.
For instance, if $ m_{3/2} >$ 100 TeV, \cite{heavy} then the gravitinos
decay before BBN ameliorating the problem. Or perhaps the gravitinos
are the LSP, but then one must insure that the NLSP does
not give rise to the same troubles. \cite{NLSP}

Lepton flavor violation (LFV) provides another example of tension
between SUSY and leptogenesis. For instance, the predictions for $\mu \to
e \gamma$, although model dependent, are sensitive to the mass of  the
heaviest neutrinos [$BR\sim (M_3 ln M_X/M_3)^2]$ and, in general,  to
satisfy the present bounds on $\mu \to e \gamma$ one needs $M_3 <
10^{13}$ GeV. \cite{Petcov} Thus LFV provides constrains from above on the spectrum of
heavy neutrinos, while leptogenesis provides constraints from below.
These examples suggest that seeking compatibility between SUSY and
leptogenesis provides interesting testable experimental predictions and
insights into neutrino physics.

\subsection{ Affleck-Dine Baryogenesis}

There is another interesting mechanism in the early Universe for
baryogenesis, which was  first discussed by Affleck and Dine.
\cite{AD} The Affleck-Dine scenario most naturally ensues in SUSY
theories where:
 
\noindent i. Some scalar fields carry B or L (e.g. squarks or sleptons).

\noindent ii. These scalar fields lie along flat directions of the
potential,  which are eventually lifted when one includes higher
dimensional  operators in the theory.

In these circumstances, it is possible for the scalar fields lying in
the flat direction  to acquire large initial values, $\phi =\phi_o$.
Eventually, when the expansion rate of the Universe H gets to be of
$O(m_{\phi})$, the field $\phi$ begins to oscillate  about the minimum
of the potential. These coherent oscillations of $\phi$ generate $\eta$.

It is useful to illustrate how the Affleck- Dine mechanism works by
means of a simple toy example. \cite{DK} Consider a Lagrangian involving a complex scalar field
\begin{equation}
 L_0=  -\partial_{\mu}\phi \partial^{\mu} \phi^{\dagger}
-m^2\phi\phi^{\dagger}.
\end{equation}
Clearly $L_0$ is invariant under the phase transformation $\phi \to e^{i
\alpha}\phi$, which plays the role of B in the model. It is also invariant under the discrete
symmetry $\phi \to \phi^{\dagger}$, which plays the role of CP in the model. If one adds to this Lagrangian the
perturbation
\begin{equation}
  L_1 = \epsilon \phi^3\phi^{\dagger}  +\epsilon^*\phi^{\dagger 3} \phi
\end{equation}
both B and CP are broken.

The evolution of $\phi(t) $ in the Universe is determined by
\begin{equation}
\ddot{\phi} + 3H\dot{\phi} +m^2\phi=\epsilon \phi^3 + 3
\epsilon^*\phi^{\dagger 2}\phi.
\end{equation}                    
If one neglects the perturbation on the RHS, and considers for
definitiveness the Hubble parameter appropriate for a matter dominated
Universe, $H= 2/3t$, the result is a damped oscillation for $\phi$  for
$H<< m$:
\begin{equation}
\phi=\phi_0 \frac{{\rm sin}mt}{mt}.
\end{equation}

When one includes the effect of the small perturbation $L_1$, besides
$\rm{Re}\phi \equiv  \phi_r = \phi_0 {\rm sin} mt / mt$,  one develops
also a small $\rm{Im} \phi\equiv \phi_i$, which obeys the equation:
\begin{equation}
\ddot{\phi}_i + \frac{2\dot{\phi}_i}{t} +m^2\phi_i \simeq 4\rm{Im}
\epsilon ~\phi_r^3.
\end{equation}
 The RHS of the above equation is negligible for large t, but sets the
size of $\phi_i$. One finds \cite{DK}
\begin{equation}
   \phi_i \simeq 4A\rm{Im} \epsilon ~\phi^3_0 \frac{ \rm{sin} (mt +
\delta)}{ m^3t},  
\end{equation}
where A and $\delta$ are parameters of O(1). The presence of $\phi_i$
allows a baryon number to develop
\begin{equation}
    n_B= i[\phi \frac{\partial \phi^{\dagger}}{\partial t}  -
\phi^{\dagger} \frac {\partial \phi}{\partial t}]
=\phi_r\dot{\phi_i}-\phi_i\dot{\phi_r}
\end{equation}
  and $n_B\simeq \rm{Im}\epsilon ~\phi^4_0 / m$ can be large even if
$\rm{ Im} \epsilon$ is small, provided $\phi_0$ is large.

 After this toy model discussion, I want to illustrate how the
Affleck-Dine mechanism for baryogenesis works in a more realistic
scenario. The scenario involves an interesting flat direction which
arises in the minimal supersymmetric extension of the SM, the MSSM.\cite
{MY, DRT} The scalar field in question is the product of the scalar
fields associated with the Higgs and lepton doublets: $\Phi^2 \equiv
(\phi_u L)$.

 For this field the seesaw mechanism induces a quartic term in the
superpotential
\begin{equation}
 W = [\frac{1}{\Lambda}] (\phi_u L)^2
\end{equation}
  which, in turn, gives rise to a potential
\begin{equation}
 V= [\frac{1}{\Lambda}]^2 |\Phi|^6.
\end{equation}
In the above, $\Lambda$ is a heavy scale.
In addition to this term in V, one expects SUSY breaking contribution to
the scalar potential involving $\Phi$ with the generic form:
\begin{equation}
  \delta V = m^2 |\Phi|^2 + [\frac{m_{SUSY}}{\Lambda}] (a \Phi^4 + a^*
\Phi^{\dagger 4 })
\end{equation}
where the mass $ m \sim m_{SUSY } \sim $ TeV.

The Affleck-Dine scenario ensues by assuming that during inflation the
field $\Phi$ acquires a negative mass-squared contribution $ - H^2\Phi^2$
which is balanced by V. Thus, the field $\Phi$ settles initially at $\Phi_0 \sim
[H \Lambda]^{1/2}$. After inflation $\Phi_0$ keeps decreasing and begins
to oscillate when the Hubble parameter is of order $H \sim  m$, at which
point $\Phi_0 \simeq [m \Lambda]^{1/2}$. 
At this time lepton-number is also generated, because $\Phi$ acquires an
imaginary component as a result of the quartic perturbation in $\delta
V$. One has:
\begin{equation}
\frac{\partial n_L}{\partial t} + 3H n_L= 4[\frac{m_{SUSY} }{\Lambda}]
{\rm Im }(a \Phi_0^4).
\end{equation}
The generated lepton number is of the order of
\begin{equation} 
   n_L \simeq [\frac{m_{SUSY}}{ \Lambda}]\Phi_0^4~ \delta_{\rm {eff}}~
t_{\rm{osc}} \simeq  m^2 \Lambda \delta_{\rm{eff}}.
\end{equation}
In the above the effective CP-violating phase $\delta_{\rm{eff}}= {\rm
sin}( {\rm arg}~ a+ 4{\rm arg} \Phi_0)$ and we have used that
$t_{\rm{osc}} \sim 1/m$.

Taking into account the entropy that is produced during inflaton decay
\cite{DRT} 
\begin{equation}
 s \sim  \frac{m^2 M_G^2}{ T_R},
\end{equation}
where $M_G$ is the reduced Planck mass $ M_G = M_P/\sqrt{8\pi} \simeq  2.4 \times 10^{18}$
GeV, gives
\begin{equation}
\eta_L= 7\frac{n_L}{s} \simeq  7\frac{\Lambda T_R }{ M_G^2}
\delta_{\rm{eff}},
\end{equation} 
 which for $T_R \sim  10^7 $ GeV , after sphaleron re-processing,
needs $\Lambda~\sim M_G$  to get  the
desired $\eta$. 

\section{Concluding Remarks}

The present situation regarding the matter-antimatter asymmetry in the
Universe is somewhat unsettled. Because $\eta$ is just one number, it is
difficult to prove any one scenario correct. From my own perspective,
perhaps the most promising scenario is leptogenesis, which connects the
magnitude of $\eta \sim 10^{-9}$  with the existence of sub-eV neutrinos.
However, the conflict of leptogenesis with "standard" SUSY should not be
ignored and, if supersymmetry is found, will have to be resolved.

From our discussion, it seems pretty clear that there is little hope to
connect $\eta$ to the CP violating CKM phase $\gamma$. For instance, in
Affleck-Dine baryogenesis the CP phase is connected with the SUSY
breaking terms in the potential, while in leptogenesis the CP phase is
connected to the heavy neutrino Yukawa couplings. These "high energy"
CP-violating phases are generally different than those measured at low
energy.

For example, in the case of leptogenesis, one can easily enumerate the
21 parameters of the lepton sector. \cite {BR}  Of these, 12 are
relevant at low energy: 3 light neutrino and 3 charged lepton masses; 3
mixing angles; and 3 CP violating phases $\alpha_1,\alpha_2$ and
$\delta$, where the last phase is the analogue of the CKM phase $\gamma$
and the other 2 phases arise because neutrinos are Majorana particles.
The other 9 parameters in the lepton sector are related to physics at
high scales. They involve the 3 heavy neutrino masses; 3 additional
mixing angles; and 3 "high scale" CP-violating phases $\phi_i$.

In a basis where the heavy neutrino mass matrix M and the charged lepton
mass matrix $m_{\ell}$ are diagonal, all 6 phases appear in the Dirac
mass matrix  $m_D$  which connects the left-handed neutrino  $SU(2)$
doublets with the right-handed neutrino singlets.The $3 \times 3$ matrix 
 $m_D$ can always
be written as a product of a unitary matrix U times a triangular matrix
Y, $ m_D = U Y$, where \cite{triangular} 
\begin{equation}
Y= \left[\begin{array}{ccc}
y_{11}&0&0\\y_{21}e^{i\phi_{21}}&y_{22}&0\\y_{31}e^{i\phi_{31}}&y_{32}e^
{i\phi_{23}}&y_{33} \end{array}\right].
\end{equation}
 The low energy CP violating phases $\alpha_1,\alpha_2 ,\delta$ in this
parametrization depend on all 6 of these phases  (3 in the unitary
matrix U and 3 in Y). On the other hand, the  CP-violating parameter
$\epsilon$ which enters in leptogenesis, because it depends on
$m_D^{\dagger} m_D$, involves only the 3 phases in Y.

Thus, without further assumptions, also in leptogenesis there is no
direct link between the matter-antimatter asymmetry and potentially
measurable CP-violating phenomena in the neutrino sector. This is a more
general problem. Without an understanding of what is the correct
underlying theory, there is no hope of identifying the phase(s) which
are responsible for $\eta$  and hence our existence, much less of
relating these phases to the {\bf ur-phase} arising from the compactification
from d=10 to d=4  dimensions, presumably responsible for the arrow of
time.

\section*{Acknowledgements}

I am extremely grateful to all the organizers of the Galapagos Summit for their wonderful hospitality. I would also like to thank Wilfried Buchm\"uller and Tsutomu Yanagida for their insights on the matter-antimatter asymmetry of the Universe. This work was supported in part by the Department of Energy under Contract No. DE-FG03-91ER40662, Task C.


\begin{thebibliography}{99}

\bibitem{Turner}  See, for example, M. S. Turner, in {\bf Intersection between Particle Physics and Cosmology}, Jerusalem Winter School for Theoretical Physics, Vol. 1, p.  99, eds. T. Piran and S. Weinberg (World Scientific, Singapore, 1986).

\bibitem{CDRG}  A. Dolgov and J. Silk, Phys. Rev. {\bf D47}, 4244 (1993); M. Y. Khlopov, M. G. Rubin, and A. S. Sakharov, Phys. Rev. {\bf D62}, 083505 (2000); A. Cohen, A. De Rujula and S. Glashow, Astrophys. J. {\bf 495}, 539 (1998).


\bibitem{WMAP} WMAP Collaboration, C. L. Bennett, {\it{et al}}, Astrophys. J. Suppl. {\bf 148}, 1 (2003); D. N. Spergel, {\it {et al}},  Astrophys. J. Suppl. {\bf 148}, 175 (2003).

\bibitem{BBN} See, for example, G. Steigman, Int. J. Mod. Phys. {\bf{E15}}, 1 (2006).

\bibitem{Sakharov} A. D. Sakharov, JETP Lett. {\bf 5}, 24 (1967).

\bibitem{ABJ} S. Adler, Phys. Rev. {\bf 177}, 2426 (1969);
J. S. Bell and R. Jackiw, Nuovo Cimento {\bf 60}, 47 (1969).


\bibitem{'tH} G. t' Hooft, Phys. Rev. Lett. {\bf 37}, 8 (1976); 
Phys. Rev. {\bf D14}, 4332 (1976).


\bibitem{KM}F. R. Klinkhamer and N. S. Manton, Phys. Rev. {\bf{D30}}, 2212 (1984).


\bibitem{KRS} V. A. Kuzmin, V. A. Rubakov and M. A. Shaposhnikov, 
Phys. Lett. {\bf{B155}}, 36 (1985).

\bibitem{Arnold} P. Arnold and L. Yaffe, Phys. Rev. {\bf{D62}}, 125014 (2000).

\bibitem{B}
D. B\"odeker, Phys. Lett. {\bf{B426}}, 351 (1998); Nucl. Phys. {\bf {B559}}, 502 (1999).


\bibitem{CPT} W. Pauli, in {\bf Niels Bohr and the Development of Physics}, ed. W. Pauli (Pergamon Press, New York, 1955); J. Schwinger, Phys. Rev. {\bf{82}}, 914 (1951); G. L\"uders, Dansk. Mat. Fys. Medd. {\bf{28}}, 5 (1954); G. L\"uders and B. Zumino, Phys. Rev. {\bf{110}}, 1450 (1958).

\bibitem{RDPS} See, for example, R. D. Peccei, in {\bf{Broken Symmetries}}, eds. L. Mathelitsch and W. Plessas, Springer Lecture Notes in Physics 521 (Springer Verlag, Berlin, 1999).

\bibitem{DLM} M. Dine, R. G. Leigh and D. A. MacIntire, Phys. Rev. Lett. {\bf{69}}, 2030 (1992).

\bibitem{CKN} K. Choi, D. B. Kaplan and A. E. Nelson, Nucl. Phys. {\bf{B391}}, 515 (1993).

\bibitem{ZKO} Y. B.  Zeldovich, I. B. Kobzarev and L. Okun, Phys. Lett. {\bf{B50}}, 340 (1974).

\bibitem{Burchat} P. Burchat, these Proceedings.

\bibitem{CKM} N. Cabibbo, Phys. Rev. Lett.  {\bf{12}}, 531 (1963); M. Kobayashi and T. Maskawa, Prog. Theor. Phys. {\bf{49}}, 652 (1973).

\bibitem{GUT} For a review see, Grand Unified Theories by S. Raby in the Particle Data Group Report, W. -M. Yao, {\it{et al}}, J. Phys. G. {\bf{33}}, 1 (2006).

\bibitem{Yoshimura}  M. Yoshimura, Phy. Rev. Lett. {\bf{41}}, 281 (1978); {\it ibid.} {\bf{ 42}}, 746(E) (1979);
  D.  Toussaint, S.~B. Treiman, F. Wilczek and A. Zee, Phys. Rev. {\bf{D19}}, 1036 (1979);
  S. Weinberg, Phys. Rev Lett. {\bf{42}}, 850 (1979);
  S. Dimopoulos and L. Susskind, Phys. Rev. {\bf{D18}}, 4500 (1978).

\bibitem{Langacker} See, for example, P. Langacker, Phys. Rept {\bf{72C}}, 185 (1981).

\bibitem{Weinberg} S. Weinberg, Phys. Rev. Lett. {\bf{43}}, 1566 (1979).

\bibitem{WZ} F. Wilczek and A. Zee,  Phys. Rev. Lett. {\bf{43}}, 1571 (1979).


\bibitem{Shap} V. A. Rubakov and M. S. Shaposhnikov, Phys. Usp. {\bf 39}, 461 (1996).

\bibitem{S} M. E. Shaposhnikov, Nucl. Phys. {\bf {B287}}, 757 (1987): {\it {ibid}}, {\bf{B299}}, 797 (1988).

\bibitem{LEP}  Particle Data Group, W. -M. Yao, {\it{et al}}, J. Phys. G. {\bf{33}}, 1 (2006).

\bibitem{jump} M. E. Shaposhnikov, JETP Lett. {\bf 44}, 465 (1986).

\bibitem {erase}  K.  Jansen, Nucl. Phys. (Proc. Supp.) {\bf B47}, 196 (1996).

\bibitem{Quiros} B. de Carlos and J. R. Espinosa, Nucl. Phys. {\bf{B503}}, 24 (1997): M. Carena, M. Quiros and C. E. M. Wagner, Phys. Lett. {\bf{B380}}, 81 (1996); M. Carena, M. Quiros, M. Seco and C. E. M. Wagner, Nucl. Phys. {\bf{B650}}, 24 (2003); K. Kainulainen, T. Prokopec, M. G. Schmidt and S. Weinstock, JHEP {\bf{0106}}, 031 (2001).

\bibitem{singlet} J. Kang, P. Langacker, T. Li and T. Liu, Phys. Rev. Lett. {\bf{94}}, 061801 (2005); S. J. Huber, T. Konstandin, T. Prokopec and M. G. Schmidt, hep-ph/0606298.

\bibitem{FHS} L. Fromme, S. J. Huber and M. Seniuch,  hep-ph/0605242.
     
\bibitem{FY} M. Fukugita and T. Yanagida, Phys. Lett. {\bf B174}, 45 (1986).

\bibitem{BPY} For an extensive review of this scenario, see for example, W. Buchm\"uller, R. D. Peccei and T. Yanagida, Ann. Rev. Nucl. Part. Sci. {\bf{55}}, 311 (2005).
         
\bibitem{HT}  J. A. Harvey and  M. S. Turner, Phys. Rev. {\bf{D42}}, 3344 (1990); R. N. Mohapatra and X. Zhang, Phys. Rev. {\bf{D45}}, 2699 (1992); S. Yu. Khlebnikov and  M. E. Shaposhnikov, Nucl. Phys. {\bf{B308}}, 885 (1988).


\bibitem{seesaw} 
T. Yanagida, in {\bf Proc. of the Workshop on ''The Unified Theory and the
   Baryon Number in the Universe''}, Tsukuba, Japan, Feb. 13-14, 1979, p. 95, 
eds. O.~Sawada and S.~Sugamoto, (KEK Report KEK-79-18, 1979, Tsukuba);
Progr. Theor. Phys. {\bf 64}, 1103 (1980); 
P. Ramond, in {\bf Talk given at the Sanibel Symposium}, Palm Coast, Fla.,
Feb. 25-Mar. 2, 1979, preprint CALT-68-709. See also
S. Glashow, in {\bf Proc. of the Cargese Summer Institute
 on ''Quarks and Leptons''}, Cargese, July 9-29, 1979, p. 707 eds. M. Levy, {\it{et al}} (Plenum, New York, 1980).

\bibitem{BDP} W. Buchm\"uller, P. Di Bari and M. Pl\"umacher, Ann. Phys. {\bf 315}, 303 (2005).

\bibitem{DI} S. Davidson and A. Ibarra, Phys. Lett. {\bf{B535}}, 25 (2002).

\bibitem{BDP2} W. Buchm\"uller, P. Di Bari and M. Pl\"umacher, Nucl. Phys. {\bf{B643}}, 367 (2002); Phys. Lett. {\bf{B547}}, 128 (2002).

\bibitem{Kawasaki} M. Kawasaki, K. Kohri and T. Moroi, Phys. Rev. {\bf{D71}}, 083502 (2005).
         
\bibitem{heavy} M. Ibe, R. Kitano, H. Murayama and T. Yanagida, Phys. Rev. {\bf{D70}}, 075012 (2004).

\bibitem{NLSP} M. Bolz, W. Buchm\"uller and M. Pl\"umacher, Phys. Lett. {\bf B443}, 209 (1998); M. Fujii, M. Ibe and T. Yanagida, Phys. Lett. {\bf B579}, 6 (2004);
J. R. Ellis, K. A. Olive, Y. Santoso and V. C. Spanos, Phys. Lett. {\bf B588} 7 (2004); 
J. L. Feng, S. Su and F. Takayama, Phys. Rev. {\bf D70}, 075019 (2004);  L. Roszkowski, R. Ruiz de Austri and  K.-Y. Choi, JHEP {\bf {0508}} 080 (2005);  
J. L. Feng, A. Rajaraman and F. Takayama, Phys. Rev. Lett. {\bf 91}, 011302-1
(2003).

\bibitem{Petcov} S. T. Petcov and T. Shindou, hep-ph/0605151 and hep-ph/0605204.

\bibitem{AD} I. Affleck and M. Dine, Nucl. Phys. B{\bf 249}, 361 (1985).

\bibitem{DK} M. Dine and A. Kusenko, Rev. Mod. Phys. {\bf{76}}, 1 (2004).

\bibitem{MY} H. Murayama and T. Yanagida, Phys. Lett. {\bf B322}, 349
(1994).

\bibitem{DRT} M. Dine, L. Randall and S. Thomas, Nucl. Phys. {\bf B458}, 291
 (1996).

\bibitem{BR}G. C. Branco, L. Lavoura and M. N. Rebelo, Phys. Lett. {\bf{B180}}, 264 (1986).

\bibitem{triangular} J. Hashida, T. Morozumi and A. Purwanto, Prog. Theor. Phys. {\bf{101}}, 379 (2000).


\end{thebibliography}
\end{document}